\documentclass[twocolumn,prb,floatfix]{revtex4-1}
\usepackage[dvipdfmx]{graphicx}
\usepackage{amsmath,amssymb}
\usepackage{bm}
\usepackage{makeidx}
\usepackage{float}
\makeindex

\begin{document}
\author{Masahiro Nakata}\altaffiliation{Present affiliation : MITSUBISHI ELECTRIC CORPORATION}
\author{Daisuke Ogura}
\author{Hidetomo Usui}
\author{Kazuhiko Kuroki}\email{Corresponding author : kuroki@phys.sci.osaka-u.ac.jp}
\affiliation{Department of Physics, 
Osaka University, 1-1 Machikaneyama, 
Toyonaka, Osaka 560-0043, Japan}
\title{Finite energy spin fluctuation as a pairing glue in systems with coexisting electron and hole bands}
\begin{abstract}
We study, within the fluctuation exchange approximation,  the spin-fluctuation-mediated superconductivity in Hubbard-type models possessing electron and hole bands, and compare them with a model on a square lattice with a large Fermi surface. In the square lattice model, superconductivity is more enhanced for better nesting for a fixed band filling. By contrast, in the models with electron and hole bands, superconductivity is optimized when the Fermi surface nesting is degraded to some extent, where finite energy spin fluctuation around the nesting vector develops.  The difference lies in the robustness of the nesting vector, namely, in models with electron and hole bands, the wave vector at which the spin susceptibility is maximized is fixed even when the nesting is degraded, whereas when the Fermi surface is large, the nesting vector varies with the deformation of the Fermi surface. We also discuss the possibility of realizing in actual materials the bilayer Hubbard model, which is a simple model with electron and hole bands, and is expected to have a very high $T_c$. 
\end{abstract}
\maketitle

\section{Introduction}
In the early days of the theoretical studies of the iron-based superconductors, the nesting between electron and hole Fermi surfaces was considered to be important for the occurrence of superconductivity. The intraorbital repulsion $U$, combined with the nesting, induces spin fluctuation, which in turn acts as a repulsive pairing interaction around the nesting vector. A repulsive pairing interaction with a certain wave vector $\bm{Q}$ generally tends to induce an unconventional superconducting gap, in which the sign changes across $\bm{Q}$. In the case of the iron-based superconductors, this mechanism leads to the so-called $s\pm$-wave pairing state, where the gap sign is reversed between the electron and the hole Fermi surfaces\cite{Hirschfeldrev,KKrev1,Chubukovrev,KKrev2,MazinSingh,KKPRL}. However, later experiments have suggested that high $T_c$ is attained  when the nesting is degraded, 
or even in the absence of the 
nesting\cite{Iimura,KFe2Se2,KFe2Se2ARPES,STO,STO2,Ding,LiOH}.
In a previous study, two of the present authors pointed out that superconductivity can be enhanced even when the nesting is not good if the magnitude of the hopping integrals matches well the inverse Fourier transformation of the gap function from momentum space to real space\cite{Suzuki}. The $s\pm$-wave pairing state corresponds to a next nearest neighbor pairing in real space, which goes hand-in-hand with the fact that the next nearest neighbor hopping is actually the dominant hopping in iron-based superconductors with high $T_c$. The importance of the real space picture implies that the states away from the Fermi level also plays an important role, since the inverse Fourier transformation involves all the states within the Brillouin zone including the states away from the Fermi level. In fact, after the observation of missing hole Fermi surface in some of the iron-based superconductors\cite{KFe2Se2,KFe2Se2ARPES,STO,STO2,LiOH,Ding,Sunagawa}, more focus has been paid on bands referred to as the ``incipient band'', which lies below, but close to the Fermi level\cite{DHLee,Hirschfeld,Hirschfeldrev,YBang,YBang2,Ding}. 

Given this background, here  we analyze two Hubbard-type models with coexisting electron and hole bands within the fluctuation exchange (FLEX) approximation\cite{Bickers,Dahm}, and compare them with the single band Hubbard model with a single large Fermi surface. We show that in models with electron and hole bands with a fixed band filling, superconductivity is optimized when the parameter values are such that the Fermi surface nesting is degraded to some extent. There, the finite energy spin fluctuation, which originates from the states away from the Fermi level, develops and acts as an effective pairing glue. This is in contrast to the case of the single band model, where better nesting gives higher $T_c$ for a fixed band filling. Owing to this difference, in an ideal situation with electron and hole bands,  $T_c$ can be much higher than in the case with a single Fermi surface.

\section{The models}
\begin{figure}
\includegraphics[width=7.5cm]{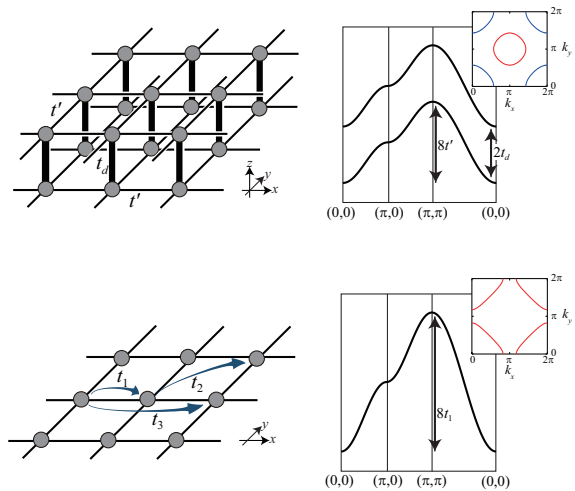}
\caption{Upper panels : the bilayer model, lower panels : the single band model on a square lattice.  Left: the lattice structure and the hoppings, right: typical band structure and Fermi surface.\label{fig1}}
\end{figure}

 As an actual superconductor with electron and hole bands, we consider the iron-based superconductor. Although the calculation results do not qualitatively depend on the material we consider, we adopt the hydrogen doped LaFeAsO as a typical example, whose five orbital model was constructed in ref.\onlinecite{Suzuki}. The electron doping rate taken here is $25 \%$. As a simpler model possessing similar kind of disconnected Fermi surfaces, we also consider a bilayer lattice system, where the bonding and the antibonding bands form electron and hole Fermi surfaces near half-filling (when the number of electrons is close to the number of sites)\cite{Bulut,KA,Scalettar,Hanke,Santos,Mazin,Kancharla,Bouadim,Fabrizio,Zhai,Maier}. The model and its band structure are shown in Fig.\ref{fig1}. For this latter model, we fix the band filling at $n=0.9$, where $n=1$ corresponds to half-filling.

In both of these systems, the nesting of the Fermi surface can be controlled by certain parameters for a fixed band filling. In the iron-based superconductor, the Fe-As-Fe bond angle $\alpha$ controls the nesting of the Fermi surface originating from the $d_{xy}$ orbital\cite{KurokiironPRB,Miyake,Usui,Andersen}.  Namely,  when the bond angle is large, the hole Fermi surface around the Brillouin zone corner $(\pi,\pi)$ is missing,  but as the angle $\alpha$ becomes smaller, the Fermi surface appears and grows larger,  so that the nesting between the electron and hole Fermi surfaces gets better. We refer to Fig.1 of ref.\onlinecite{Suzuki} for the variation of the Fermi surface with the change in the bond angle. In the bilayer model, the ratio between the intralayer and the interlayer hoppings, $t'$ and $t_d$, respectively, controls the Fermi surface nesting for non-half-filled cases. When $t'/t_d$ is reduced,  the overlap between the bonding and the antibonding bands becomes smaller to degrade the nesting, and eventually one of the Fermi surfaces disappears. Here $t_d$ is fixed at 1.4 as in ref.\onlinecite{KA}, and hence $t_d/1.4$ is the unit of the energy.

As an example of systems with a large Fermi surface, we consider a single band Hubbard model on a square lattice, which is often considered as an effective model for the high $T_c$ cuprates. As shown in the lower panel of Fig.\ref{fig1}, we consider hoppings up to third nearest neighbors, $t_1$ (fixed at $-1$ here, so that  $|t_1|$ is the unit of the energy), $t_2$, and $t_3$, with $t_2=-t_3$ for simplicity (this relation is roughly satisfied in the cuprates). For a fixed band filling close to half-filling, the Fermi surface nesting becomes degraded when $t_2(=-t_3)$ is increased. For the single band model, we fix the band filling at $n=0.85$ (corresponds to 15\% hole doping in the cuprates).

On top of these tightbinding models, we take into account the on-site electron-electron interactions, that is, the on-site Hubbard $U$ for the bilayer and the square-lattice models, and the multiorbital interactions for the iron-based superconductor in the form
\begin{eqnarray}
H_1=\sum_{i}\left[\sum_{\mu}U_{\mu}n_{i\mu\uparrow}n_{i\mu\downarrow}+\sum_{\mu>\nu}\sum_{\sigma\sigma'}U'_{\mu\nu}n_{i\mu\sigma}n_{i\mu\sigma '} \right. \nonumber \\ \left.
-\sum_{\mu\neq\nu}J_{\mu\nu}\bm{S}_{i\mu}\cdot\bm{S}_{i\nu}+\sum_{\mu\neq\nu}J'_{\mu\nu}c_{i\uparrow}^{\mu\dag}c_{i\downarrow}^{\mu\dag}c_{i\downarrow}^{\nu}c_{i\uparrow}^{\nu}\right] \nonumber
\end{eqnarray}
in notations adopted in ref.\onlinecite{KurokiironPRB}.  
We apply the FLEX approximation to obtain the renormalized Green's function, the spin and charge susceptibilities, which are plugged into the Eliashberg equation to study superconductivity. The  values of the electron-electron interactions for the model of the iron-based superconductor are $U=1.3$ eV, $U'=2U/3$ and $J=J'=U/6$. $U=6$ and $U=8$ for the square lattice and the bilayer model, respectively, in units adopted in each model.
The temperature is $T=0.01$eV for the model of the iron-based superconductor, and $T=0.1$ and $T=0.03$ for the bilayer and the square lattice model in units taken for each model. The number of $k$-point meshes is $32\times 32$ for the square lattice, and $64\times 64$ for the iron-based superconductor and the bilayer model. The number of Matsubara frequencies is 2048 for the square lattice and the iron-based superconductor, and 4096 for the bilayer. 
 
We focus on the imaginary part of the dynamical spin susceptibility $\chi(\bm{q},\omega)$, which is experimentally observed in NMR and neutron scattering experiments as a measure for the strength of the spin fluctuation in momentum and energy space. $\chi(\bm{q},\omega)$ is obtained by Pad\'{e} analytical continuation of the FLEX spin susceptibility obtained within the Matsubara formalism.
For the model of the iron-based superconductor, we obtain the susceptibility in the orbital representation and take the diagonal element $\chi^{\mu\mu\mu\mu}(\bm{q},i\omega)$ ($\mu$ denotes the orbitals, see ref.\onlinecite{KurokiironPRB}) with $\mu=d_{xy}$ orbital, since it is the $d_{xy}$ orbital portion of the Fermi surface that is controlled by the bond angle. As for the bilayer model, $\chi$ is a $2\times 2$ matrix\cite{KA}, and we take the trace of this matrix as $\chi(\bm{q},i\omega)$.  Since the spin fluctuation around the nesting vector acts as the Cooper pairing glue, we sum up $\chi(\bm{q},\omega)$ (in the area of $(1/8)^2$  of the entire Brillouin zone) around a certain wave vector $\bm{Q}$, and this is defined as $\Gamma(\omega)$. 
\begin{equation}
\sum_{\bm{q}\sim\bm{Q}}\mathrm{Im\,}\chi(\bm{q},\omega)\equiv \mathrm{Im\,}\Gamma(\omega)
\end{equation}
$\bm{Q}=(\pi,\pi)$ for the bilayer and square-lattice models (see Fig.\ref{fig1}), and $\bm{Q}=(\pi,0)$ for the iron-based superconductor (see, e.g., Fig.1 of ref.\onlinecite{Suzuki}). As for superconductivity, the Green's function is plugged into the linearized Eliashberg equation, whose eigenvalue $\lambda$ reaches unity at $T=T_c$ of superconductivity.  Here we calculate $\lambda$ at a fixed temperature for each system to measure how close the system is to superconducting transition.

\begin{figure}
\includegraphics[width=7.5cm]{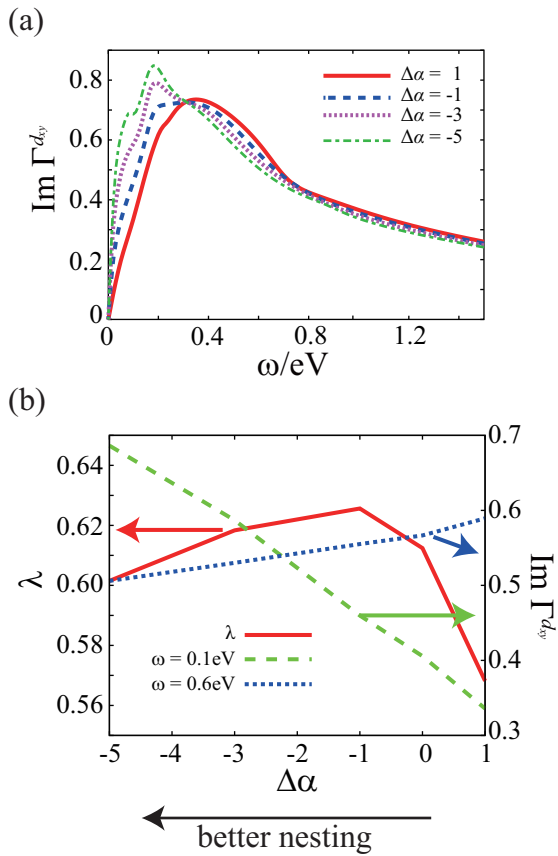}
\caption{FLEX calculation result for the five orbital model of the iron-based superconductor. (a) Im $\Gamma^{d_{xy}}$ as functions of $\omega$ for various bond angles. (b) Eigenvalue of the $s_\pm$ superconducting state (left axis) and Im $\Gamma^{d_{xy}}$ (right axis) for low and high $\omega$ values, plotted against the bond angle. \label{fig2}}
\end{figure}

\begin{figure}
\includegraphics[width=7.5cm]{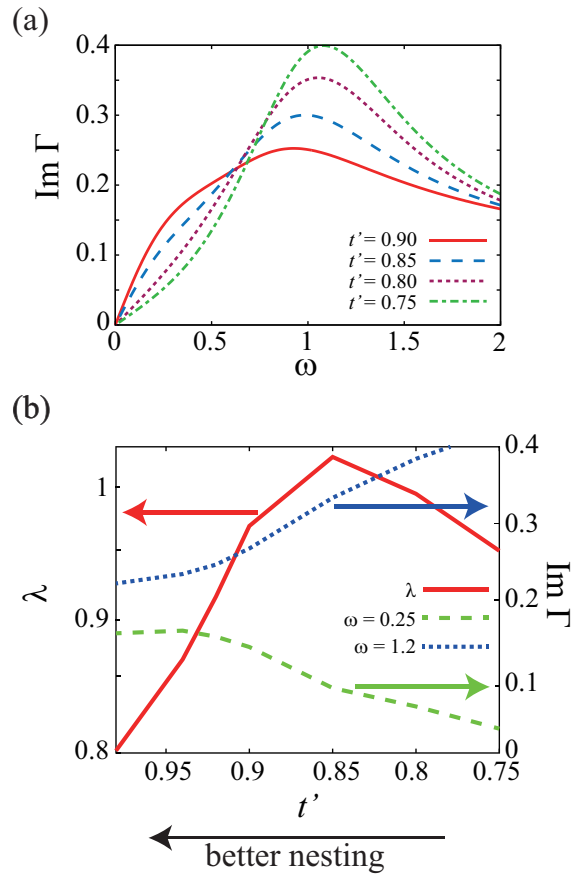}
\caption{Plots similar to Fig.\ref{fig2} for the bilayer model. Here, the parameter that controls the nesting is the intralayer hopping $t'$. \label{fig3}}
\end{figure}

\begin{figure}
\includegraphics[width=7.5cm]{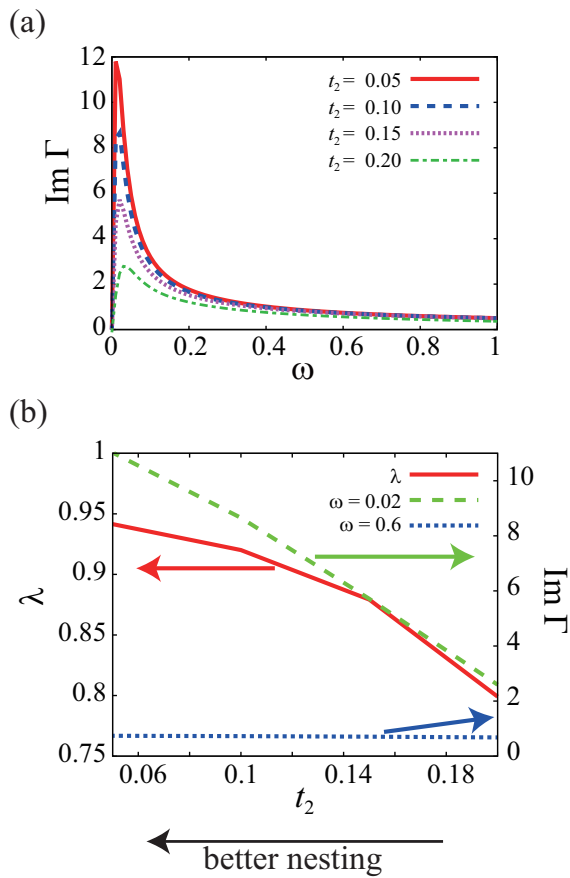}
\caption{Plots similar to Fig.\ref{fig2} for the square lattice single band model.Here, the parameter that controls the nesting is the distant hoppings $t_2=(-t_3)$.\label{fig4}}
\end{figure}

\begin{figure}
\includegraphics[width=7.0cm]{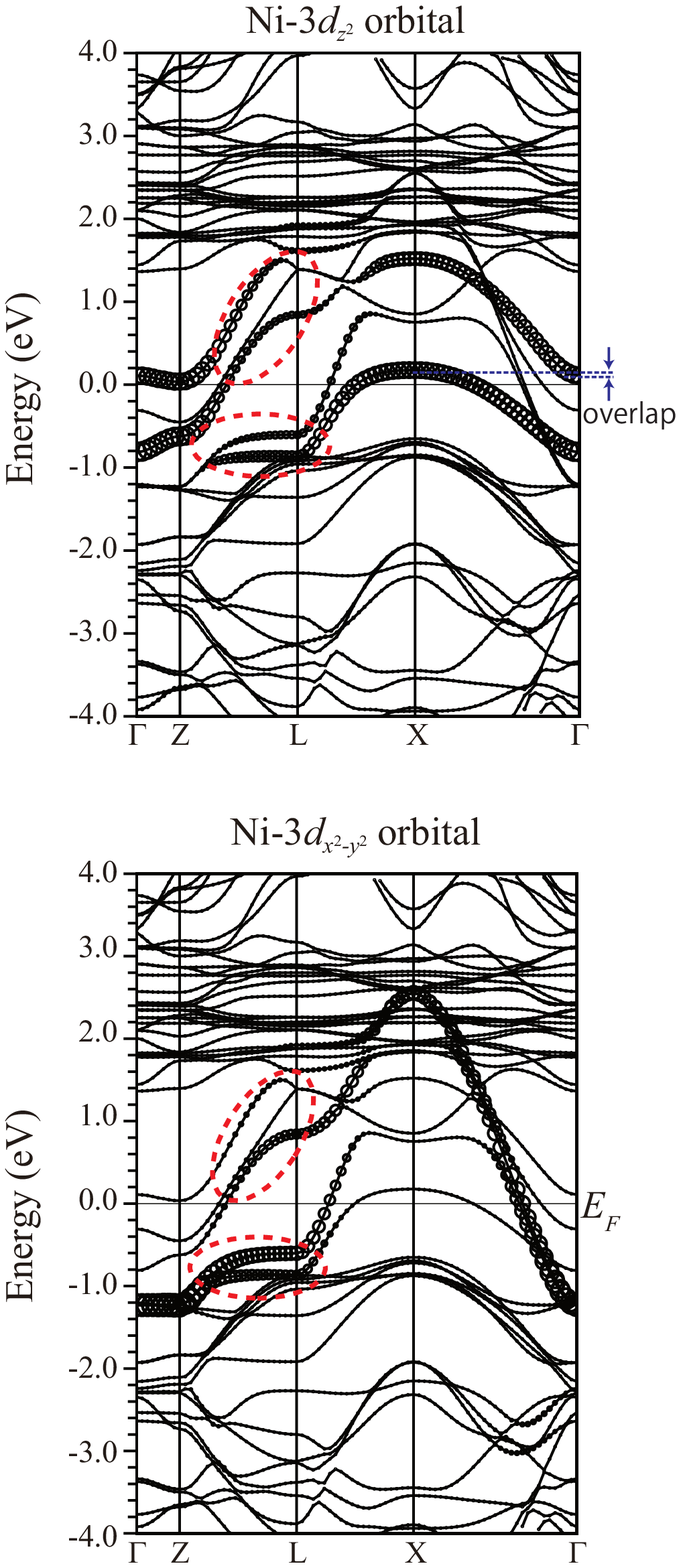}
\caption{Band calculation result for La$_3$Ni$_2$O$_7$. The thickness of the lines in the upper (lower) panel depict the $d_{3z^2-r^2}$ ($d_{x^2-y^2}$) weight. The $d_{3z^2-r^2}$ and the $d_{x^2-y^2}$ orbitals strongly hybridize in the portion encircled by the dashed line.\label{fig5}}
\end{figure}

\section{RESULTS}
We start with the iron-based superconductors. In Fig.\ref{fig2}(a), we plot Im$\Gamma(\omega)$  as functions of $\omega$ for various hypothetical values of the bond angle. Here, the bond angle is presented as the difference $\Delta\alpha$ from the bond angle of LaFeAsO$_{1-x}$H$_x$ observed experimentally\cite{Suzuki}.  When the bond angle is small and hence the Fermi surface nesting is good, the spin fluctuation around the wave vector $(\pi,0)/(0,\pi)$ is large in the small $\omega$ regime, but when the bond angle becomes large and the nesting is degraded, the low energy spin fluctuation loses weight, and the fluctuation weight moves towards higher energy regime.
In Fig.\ref{fig2}(b),  the eigenvalue of the Eliashberg equation is plotted as a function of the Fe-As-Fe bond angle, along with the value of Im$\Gamma$ at certain high and low energies. As already expected from Fig.\ref{fig2}(a), the high and low energy spin fluctuations have the opposite tendency with respect to the Fermi surface nesting. Superconductivity is optimized at a certain bond angle,  which can be considered as a consequence of the spin fluctuation  lying in a finite "sweet spot" energy regime for superconductivity.
In other words,  superconductivity is optimized for moderate Fermi surface nesting, and neither too good nor too ill-conditioned nesting is favorable for superconductivity. Since $\lambda$ is positively correlated with $T_c$, the present result explains the experimental observation that there is an optimal Fe-As-Fe bond angle for the iron-based superconductors\cite{CHLee}. Also, as was pointed out by two of the present authors in ref.\onlinecite{Arai}, the above picture explains the experimental (NMR) observation found in some of the iron-based superconductors that the strength of the low energy spin fluctuation is not positively correlated with $T_c$\cite{Ishida,Mukuda1,Mukuda2}.

In Fig.\ref{fig3}, we present similar calculation results for the bilayer model.  When $t'$ is small so that the Fermi surface nesting is good,  the spin fluctuation has large weight in the low energy regime, whereas when $t'$ is large and the nesting is degraded, the spin fluctuation weight moves towards higher energy regime. Consequently, superconductivity is optimized at a certain $t'$ that gives moderate nesting. These results suggest that the two models with disconnected Fermi surfaces share a common feature : superconductivity is optimized for somewhat degraded Fermi surface nesting, for which the spin fluctuation around the nesting vector has large weight in a finite sweet spot energy regime for superconductivity\cite{commentFS}. 

We now move on to the square lattice  for comparison. In this case,  the spin fluctuation around the wave vector $(\pi,\pi)$ is monotonically reduced in the entire energy range as $t_2(=-t_3)$ is increased to degrade the Fermi surface nesting, as shown in Fig.\ref{fig4}(a). Consequently, the eigenvalue of the $d$-wave superconductivity monotonically decreases with increasing $t_2$\cite{commentt'}. This is in sharp contrast to the cases when the Fermi surface consists of disconnected pockets.

\section{DISCUSSION}
The origin of the above results can be understood as follows. In the case of  systems with disconnected Fermi surfaces, the nesting vector, i.e., the wave vector where the spin susceptibility is maximized, is basically unchanged from its original position ($(\pi,\pi)$ for the bilayer model and $(\pi,0)/(0,\pi)$ for the iron-based superconductor)  even when the nesting is degraded. This is because the Fermi surface is small so that there are no other candidates for the nesting vector (to be precise, this is not completely correct for the iron-based superconductor; we will come back to this point later). On the other hand, when the Fermi surface is large as in the present single band model, there will appear other candidates for the nesting vector when the shape of the Fermi surface is deformed. Therefore, the weight of the spin fluctuation will be spread and transferred to various wave vector positions. This is unfavorable for $d$-wave superconductivity since its gap function has to change its sign across the wave vector at which the spin fluctuation is maximized; if a wave vector at which the spin fluctuation has significant weight connects the portion of the Fermi surface with the same sign of the superconducting gap, that will give a negative contribution to superconductivity.

In the case of the disconnected Fermi surfaces, the weight of the spin fluctuation is transferred mainly in the energy direction without basically changing the wave vector  when the nesting is degraded, so that there is a large potential of the optimized $T_c$ being higher than in the case with a large Fermi surface. In fact, previous studies have shown that the bilayer model exhibits extremely high $T_c$ compared to the single band Hubbard model\cite{KA,Maier}. This is also reproduced in the present study, as seen in Fig.\ref{fig3}(b), where the eigenvalue of the Eliashberg equation $\lambda$ exceeds unity at $t'=0.85$ and $T=0.1$, meaning that $T_c$ is larger than $0.1t'$. This $T_c$ is more than three times higher than that ($\sim 0.03t$) of the single band model\cite{commentU}. It is also worth noting that the high $T_c$ in the bilayer model is realized despite the fact that the spin fluctuation is not so strong, as seen by comparing the absolute value of Im$\Gamma$ (Fig.\ref{fig3}(a)) to that of the single band model (Fig.\ref{fig4}(a)). This implies that the high $T_c$ superconductivity in the bilayer model can take place without closely competing with magnetism, once the appropriate parameter values are realized.

$T_c$ of the bilayer model is also much higher than that of the  model of the iron-based superconductors, although the latter also has disconnected Fermi surfaces\cite{commentU}. There are actually two reasons for this. (i) The multiorbital nature degrades the effect of the spin-fluctuation-mediated pairing glue ; the spin fluctuation can be most effective as a pairing glue when only one up and one down spin electrons can occupy the same site, that is, in the case of single orbital systems. (ii) When the bond angle becomes so large that the $d_{xy}$ Fermi surface around $(\pi,\pi)$ is completely lost, spin fluctuation develops around another nesting vector $(\pi/2,\pi)/(\pi,\pi/2)$ between the two electron Fermi surfaces\cite{KurokiironPRB,Graser}. Therefore, the nesting vector is not as robust as in the bilayer model. 

The bilayer model is, at present, a toy model which does not correspond to an actual  material, but the above discussion suggests that if a corresponding material is found, it may have $T_c$ even higher than that of the cuprate superconductors. It is thus tempting to search for materials that actually corresponds to the present bilayer model. An important feature of this model is that the interlayer hopping $t_d$ is moderately (about two times) larger than the intralayer one $t'$. Hence, candidates may be found in materials with nearly half-filled $d_{3z^2-r^2}$ orbital, rather than the $d_{x^2-y^2}$ orbital that plays the main role in the cuprates. As a possible candidate toward this direction, we focus on a particular existing Ruddlesden-Poppers compound, La$_3$Ni$_2$O$_7$, which has a bilayer structure of NiO$_2$ planes. The $3d^{7.5}$ configuration in the nickelate, rather than the $3d^9$ in the cuprates, makes the $d_{3z^2-r^2}$ band close to half filling. The material is non-superconducting, and previous studies have suggested that the $d_{3z^2-r^2}$ orbital is in the half-filled Mott insulating state, while the overlapping quarterfilled band originating from the $d_{x^2-y^2}$ orbital is metallic\cite{Kobayashi_JPSJ}. Nonetheless, we perform a band calculation\cite{Pickett} to see whether a band structure similar to that of the bilayer model could emerge, were it not for the Mott insulating state. We carry out the band calculation using the Wien2k package\cite{wien},  assuming an ideal, undistorted lattice structure and a  paramagnetic phase, and adopting the experimentally determined lattice structure\cite{Zhang}. We set $RK_{max}=6.5$ and employ 512 $k$-meshes.

The obtained band structure is shown in Fig.\ref{fig5}.  If we focus on the bands originating from the $d_{3z^2-r^2}$ orbital, we can see that there indeed exist  bonding and antibonding bands, similar to those of the bilayer model. However, the overlap between the two bands is barely present, meaning that $t_d$ is about $4t'$ (see Fig.\ref{fig1}), which is somewhat large from the viewpoint of the present calculation results. If the layer-layer distance can somehow be enlarged (and, of course, the orbital selective Mott insulating state of the $d_{3z^2-r^2}$ orbital can be circumvented), the situation will be closer to that of the bilayer model. Another point that should be noticed is the $d_{x^2-y^2}$ orbital contribution; the $d_{x^2-y^2}$ band strongly overlaps with the $d_{3z^2-r^2}$ band, and there is a $d_{3z^2-r^2}$-$d_{x^2-y^2}$ hybridization around the portion encircled by the dashed lines, which deforms the $d_{3z^2-r^2}$ band from the ideal shape shown in Fig.\ref{fig1}. This hybridization may degrade superconductivity, as was shown in the case of the cuprates ; there, the situation is the opposite, namely, the main band is the $d_{x^2-y^2}$ band, and the hybridization with the $d_{3z^2-r^2}$ orbital degrades the superconductivity\cite{SakakibaraPRL,SakakibaraPRB}. In the present case, the $d_{x^2-y^2}$ band may have to be lifted to higher energies, using, e.g., certain modification of the crystal field,  to reduce the hybridization effect. Further study along this line is in progress.

\section{CONCLUSION}
We have studied the spin-fluctuation-mediated superconductivity in Hubbard type models possessing electron and hole bands, and compared them with the Hubbard model on a square lattice with a large Fermi surface. In the former models, superconductivity is optimized when the Fermi surface nesting is degraded to some extent, and finite energy spin fluctuations around the nesting vector develops. This is contrast to the case of the square lattice model, where superconductivity is more enhanced for better nesting. The difference lies in the robustness of the nesting vector, namely, in models with electron and hole bands, the wave vector at which the spin susceptibility is maximized is fixed even when the nesting is degraded, whereas when the Fermi surface is large, the nesting vector varies with the deformation of the Fermi surface. 

As seen in the calculation result for the bilayer model, the large enhancement of the finite energy spin fluctuation around the robust nesting vector can give rise to an extremely high $T_c$, although such a situation is realized only in a narrow ``sweet spot'' regime in the parameter space. Our band calculation result shows that such a situation might be realized by modifying the lattice structure and the constituting elements of existing bilayer materials.  

\section*{Acknowledgment}
We thank Hiroshi Eisaki for pointing out the possible relevance of La$_3$Ni$_2$O$_7$ to the bilayer model. We also thank Hideo Aoki and Kenji Kawashima for discussions on the realization of the bilayer model in actual materials. We also appreciate Katsuhiro Suzuki for the assistance with the multiorbital FLEX code. Part of the calculation was performed using supercomputers at the Supercomputer Center, Insitute for Solid State Physics, The University of Tokyo. This study was supported by  Grants-in-Aid for Scientific Research (A) (No.JP26247057) and Grants-in-Aid for Scientific Research (B) (No.JP16H04338) 
from the Japan Society for the Promotion of Science. D.O. acknowledges support from Grant-in-Aid for JSPS Research Fellow Grant No. 16J01021.


\begin{thebibliography}{99}
\bibitem{commentFS} For the parameter values that optimize  superconductivity in these systems, the hole and the electron Fermi surfaces both exist in the non-interacting band structure, i.e., the band structure for $U=0$. However, in the actual renormalized band structure, the presence/absence of one of the Fermi surfaces becomes a subtle problem, as can be roughly seen from the FLEX calculation result of the dressed Green function at the lowest Matsubara frequency. Here we will not go into this problem since a precise evaluation requires highly accurate analytical continuation.
\bibitem{commentU} Although we take somewhat different values of $U$ between the bilayer and the square lattice models, this does not affect the conclusion here since $T_c$ is not strongly dependent on the value of $U$ around these values. Also, the intraorbital $U=1.3$eV of the model of the iron-based model is of the same order with those in other models when measured in units of a typical hopping integral in the iron-based superconductors, i.e. order of 0.1eV. A similar comment applies also to the difference in the doping rate among the three models.
\end{thebibliography}
\end{document}